\documentclass[aps,prl,twocolumn,showpacs,superscriptaddress,groupedaddress]{revtex4}  

\usepackage{graphicx}  
\usepackage{dcolumn}   
\usepackage{bm}        
\usepackage{amssymb}   
\usepackage{natbib}
\usepackage{hyperref}
\newcommand{\apjl}{\textrm{Astrophys. J. Lett.}}
\newcommand{\mnras}{\textrm{Mon. Not. Roy. Astro. Soc.}}
\newcommand{\aap}{\textrm{Astron. \& Astrophys.}}
\newcommand{\physrep}{\textrm{Phys. Rep.}}

\newcommand{\pd}[2]{\frac{\partial #1}{\partial #2}}

\newcommand{\thd}[3]{\biggl(\pd{#1}{#2}\biggr)_{#3}}

\newcommand{\be}{\begin{equation}}
\newcommand{\ee}{\end{equation}}
\newcommand{\bea}{\begin{eqnarray}}
\newcommand{\eea}{\end{eqnarray}}
\newcommand{\ba}{\begin{array}}
\newcommand{\ea}{\end{array}}

\long\def\symbolfootnote[#1]#2{\begingroup%
\def\thefootnote{\fnsymbol{footnote}}\footnote[#1]{#2}\endgroup}

\begin{document}
\title{Proto-Neutron Star Cooling with Convection: The Effect of the Symmetry Energy}

\author{L.F.~Roberts} \affiliation{Department of Astronomy and Astrophysics, University of California, Santa Cruz, CA 95064 USA}
\author{G.~Shen} \affiliation{Theoretical Division, Los Alamos National Laboratory, Los Alamos, NM 87544}
\author{V.~Cirigliano} \affiliation{Theoretical Division, Los Alamos National Laboratory, Los Alamos, NM 87544}
\author{J.A.~Pons} \affiliation{Departament de F\'{\i}sica Aplicada, Universitat d'Alacant, Ap. Correus 99, 03080 Alacant, Spain}
\author{S.~Reddy} \affiliation{Theoretical Division, Los Alamos National Laboratory, Los Alamos, NM 87544}
\affiliation{Institute for Nuclear Theory, University of Washington, Seattle, Washington 98195}
\author{S.E.~Woosley} \affiliation{Department of Astronomy and Astrophysics, University of California, Santa Cruz, CA 95064 USA}

\begin{abstract} 
We model neutrino emission from a newly born neutron star subsequent to a supernova explosion to study its sensitivity to the equation of state, neutrino opacities, and convective instabilities at high baryon density.  We find the time period and spatial extent over which convection operates is sensitive to the behavior of the nuclear symmetry energy at and above nuclear density.  When convection ends within the proto-neutron star, there is a break in the predicted neutrino emission that may be clearly observable.  
\end{abstract} 




\maketitle 


The hot and dense proto-neutron star (PNS) born subsequent to core-collapse in a type II supernova explosion is an intense source of neutrinos of all flavors.  It emits the $3-5 \times10^{53}$ ergs of gravitational binding energy gained during collapse as neutrino radiation on a time scale of tens of seconds as it contracts, becomes increasingly neutron-rich and cools. Cooling of the PNS and the concomitant neutrino emission are driven by neutrino diffusion and convection along the lepton number and entropy gradients left behind within the PNS after core-bounce, where the matter density and temperature are in the range $\rho=2-6\times 10^{14}$ g/cm$^3$ and $T=5-40$ MeV, respectively \citep{Burrows86,Wilson88,Keil95a,Pons99,Fischer10,Huedepohl10}.  While the supernova explosion mechanism and associated fall back of material are expected to influence the neutrino emission at early time  (i.e. $t \lesssim 1 $ s post bounce) the late time neutrino signal is shaped by the properties of the PNS, such as the nuclear equation of state (EoS), neutrino opacities in dense matter, and other microphysical properties that affect the cooling timescale by influencing either neutrino diffusion or convection \citep{Keil95b,Pons99,Pons01a,Pons01b}.  

Here, we present one-dimensional hydrostatic models of PNS evolution out to late times for two EoSs.  We include approximate convective transport along with diffusive neutrino transport, both consistent with the underlying EoS.  This allows us to gauge the importance of convection and the effect of medium modifications to the neutrino opacities in dense matter to the temporal characteristics of the neutrino signal.  The basic framework for PNS evolution is similar to that described in \cite{Pons99}, except that a treatment of convection is included.  We find that the behavior of the nuclear symmetry energy (which is a measure of the energy difference between dense neutron matter and symmetric nuclear matter) at high density significantly influences convection and thereby affects the observable neutrino signal. 

Large scale convective overturn of material will directly transport energy and lepton number in the PNS and alters the gradients along which neutrinos diffuse, thereby strongly affecting the neutrino signal accompanying PNS formation.  It has long been recognized that the outer PNS mantle is unstable to convection soon after the passage of the supernova shock, due to negative entropy gradients \citep{Epstein79}.  This early period of instability beneath the neutrino spheres has been studied extensively in both one and two dimensions, with the hope that it could increase the neutrino luminosities enough to lead to a successful explosion \citep{Burrows87,Wilson88,Keil96,Mezzacappa98a,Dessart06, Buras06b}.  Although the role of convection at late times was studied in Refs.~\citep{Burrows87,Wilson88}, this work is the first attempt at exploring its connection to the underlying microphysics and the interplay between modified opacities and convection in shaping the observable neutrino signal.   

We model convection as a diffusive process described by standard time dependent mixing length theory \cite{Wilson88}. In Ref.~\cite{Buras06b} it was shown that hydrodynamic simulations in two-dimensions were well reproduced by a simple mixing scheme in one-dimension, suggesting that this is a reasonable approximation.  Further, since convection proceeds efficiently throughout our simulations, our results are not particularly sensitive to the chosen mixing length.  The linear growth rates are obtained from the standard Ledoux stability analysis as
\be
\omega^2 = -\frac{g}{\gamma_{n_B}}\left(\gamma_s \nabla\ln(s)
+ \gamma_{Y_L} \nabla \ln(Y_L) \right),
\ee
where 
\bea
\gamma_{n_B} = \thd{\ln P}{\ln n_B}{s,Y_L}
\gamma_{s} = \thd{\ln P}{\ln s}{n_B,Y_L}
\gamma_{Y_L} = \thd{\ln P}{\ln Y_L}{n_B,s}\nonumber 
\eea
and $g$, $P$, $s$, $n_B$, and $Y_L$ are the local acceleration due to gravity, the pressure, entropy per baryon, baryon number density and the fraction of leptons in dense matter, respectively. Convective instability sets in for $\omega^2>0$.  The form of the growth rate clearly implies that there will be a strong interplay between the nuclear EoS and the patterns of convection within the PNS. 

In addition, the PNS may be subject to doubly diffusive instabilities due to the lateral transport of composition and energy by neutrinos \citep{Wilson88,Bruenn96,Miralles00}. Early one-dimensional work \cite{Wilson88} suggested that neutron fingering instabilities and convection enhanced the neutrino luminosity to successfully power a neutrino driven explosion.  However, more recent two dimensional studies found no evidence of these doubly diffusive instabilities \citep{Buras06b,Dessart06}.  Because of this and the increased complexity of treating the doubly diffusive instabilities, we do not include them in our study.

\begin{figure}
\includegraphics[scale=0.3]  {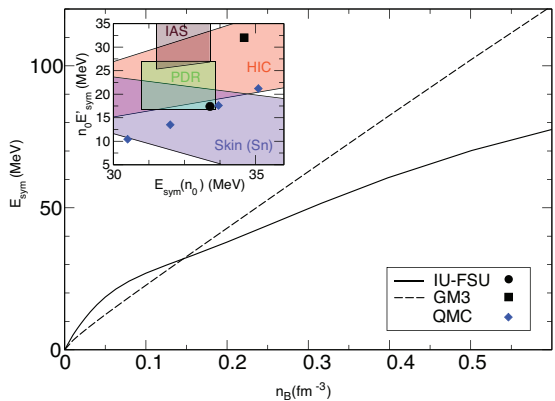}
\caption{The symmetry energy as function of density for the IU-FSU and GM3 EoSs. Inset: $n_0 E'_{\rm sym}$ versus $E_{\rm sym}$ at nuclear saturation density, for IU-FSU (circle), GM3 (square), and QMC (diamonds).  The shaded regions correspond to various experimental constraints taken from Ref.~\cite{Tsang11}.}
\label{fig:sym_energy}
\end{figure} 

The EoS and neutrino interaction rates are modeled using a relativistic mean field (RMF) model of nuclear interactions. We adopt a non-linear generalization of the original Walecka model described in \cite{Fattoyev10}. Here, the nucleon-nucleon interaction energy is calculated in the mean field approximation using effective interactions, which are tuned to reproduce gross observed properties of nuclei and empirical properties of symmetric nuclear matter at saturation density. Although these empirical constraints provide valuable guidance to constrain aspects of the symmetric nuclear EoS at nuclear densities, the experimental constraints on the properties of neutron-rich matter are relatively weak. The difference between the energy of symmetric matter (equal numbers of neutrons and protons) and pure neutron matter is called the symmetry energy, $E_{\rm sym}(n_B)$, and is defined by $E(n_B,x_p)= E(n_B,x_p=1/2) + E_{\rm sym}(n_B) \delta^2 + \cdots$.  Here, $\delta =(1-2 x_p)$ and $E(n_B,x_p)$ is the energy per particle of uniform matter composed of neutrons and protons with total baryon density $n_B$ and proton fraction $x_p$.  In charge neutral matter $x_p=Y_e$ where $Y_e$ is the electron fraction.  Various experimental probes of the nuclear symmetry energy and its density dependence in nuclei and heavy-ion collisions are actively being pursued in terrestrial experiments, but are yet to yield strong constraints. These constraints are shown in the inset in Fig.~\ref{fig:sym_energy} and are discussed in Refs.~\cite{Fattoyev10,Tsang11}.  Quantum Monte Carlo (QMC) results are also shown in the inset in Fig.~\ref{fig:sym_energy}.  The linear correlation between $E_{\rm sym}$ and $E'_{\rm sym}$ in the QMC results is obtained by varying values of the poorly known three-neutron interaction \cite{Gandolfi11}. 

Recent work has shown that the derivative of the symmetry energy with respect to density, denoted as $E'_{\rm sym}=\partial E_{\rm sym}/\partial n_B$, plays a crucial role both in the terrestrial context where it affects the neutron density distribution in neutron-rich nuclei and in astrophysics where it affects the structure and thermal evolution of neutron stars (for a recent review see Ref.~\cite{Steiner05}).  The pressure of neutron matter at saturation density,  $P_{\rm neutron}(n_0)= n_0^2 E'_{\rm sym}$, influences the radii of cold neutron stars \cite{Lattimer01}. In neutron-rich nuclei, the neutron-skin thickness is also sensitive to $E'_{\rm sym}(\rho_0)$, so that there exists a linear correlation between the neutron-skin thickness and neutron star radius \cite{Horowitz01}.

To study the sensitivity of PNS evolution to the nuclear symmetry energy we employ two RMF models with different predictions for $E'_{\rm sym}(\rho_0)$.  The first EoS is the IU-FSU EoS taken from \cite{Fattoyev10}, which includes a non-linear coupling between the vector and iso-vector mesons that allows the symmetry energy to be tuned at high density.  The second EoS employed is the GM3 parameter set, where non-linear coupling of the vector meson fields is neglected \citep{Glendenning91}.  The symmetry energy as a function of density is shown in Fig. \ref{fig:sym_energy} for the two EoS.   The inset in Fig.~\ref{fig:sym_energy} shows current theoretical estimates and experimental constraints on $E_{\rm sym}$ and $n_0E'_{\rm sym}$ at nuclear density. 

In the rest of this letter, we demonstrate that $E'_{\rm sym}(\rho_0)$ plays a role in stabilizing PNS convection at late times and thereby directly affects the PNS neutrino signal.  The logarithmic derivatives $\gamma_s$ and $\gamma_{n_B}$ are always positive, so that negative entropy gradients always provide a destabilizing influence.  For given entropy and lepton gradients, stability is then determined by the ratio $\gamma_{Y_L}/\gamma_s$.  The sign and magnitude of $\gamma_{Y_L}$ is strongly influenced by the density dependence of the nuclear asymmetry energy, so that negative gradients in lepton number can be either stabilizing or destabilizing and the degree to which they are stabilizing varies from EoS to EoS.  To clarify this we note that at $T=0$  and when the neutrino contribution to the pressure is small 
\be
\label{eq:dpdyl}
\thd{P}{Y_L}{n_B} \simeq n_B^{4/3} Y_e^{1/3} - 4 n_B^2 E'_{\rm sym}(1-2 Y_e),
\ee 
which is a reasonable approximation to the finite temperature result.  The first term comes from the electron contribution to the pressure, while the second term is due to nucleons and is negative since both the Fermi and interaction energies favor a symmetric state.  For high densities and low electron fractions, for realistic values of $E'_{\rm sym}$, this leads to negative $\gamma_{Y_L}$.  Therefore, a larger $E'_{\rm sym}$ leads to negative lepton gradients in the PNS providing a larger stabilizing influence.  $E'_{\rm sym}$ also partially determines the equilibrium value of $Y_e$, which can alter the value of $\gamma_{Y_L}$, but this is a smaller effect.  In our numerical PNS simulations this effect is accounted for.  In contrast, the properties of zero temperature nuclear matter should have significantly less effect on the behavior	of $\gamma_s$.  We find the variation of $\gamma_s$ to be significantly less than that of $\gamma_{Y_L}$ in the two EoSs considered here, which is consistent with expectations.

Medium modifications of the neutrino interaction rates also influence PNS evolution.  These effects have been investigated in earlier work where neutrino scattering and absorption rates on nucleons and leptons were calculated within the relativistic random phase approximation (RPA) \citep{Reddy99}. Here, the effects due to strong and electromagnetic correlations, degeneracy, and relativistic currents are included in a similar way.  Further, the effective interactions we employ in the RPA are obtained from the underlying model for the EoS to ensure that the correlation functions which determine the neutrino scattering kernels satisfy the generalized compressibility sum-rules consistent with the EoS.  Because the axial portion of the response accounts for the majority of the neutrino scattering cross-section, the presence of an axial interaction can strongly affect the total neutrino mean free path.  To take this into account, we introduce an effective short-range interaction in the spin channel through the Migdal parameter, $g'$ \citep{Kim95}.  The strength of this interaction is tuned to reproduce the spin-susceptibility of neutron matter obtained from microscopic calculations \cite{Fantoni01}.  For densities above nuclear saturation, the RPA causes a significant enhancement of the mean free path relative to the mean field approximation due to the repulsive nature of the nuclear interaction at high density.  

\begin{figure}
\includegraphics[scale=0.25]  {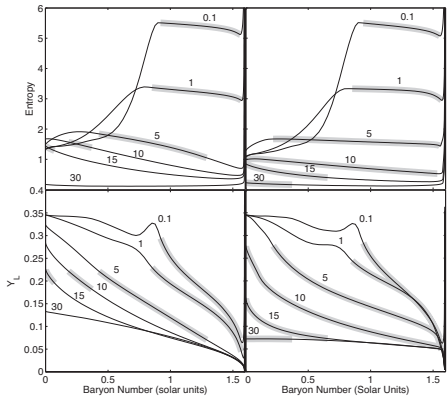}
\caption{Evolution of the entropy (top panels) and lepton fraction (bottom panels) in a $1.6 M_\odot$ rest mass PNS for the GM3 EoS (left panels) and 
the IU-FSU EoS (right panels).  The grayed regions are convectively unstable.
The labels correspond to the model times in seconds.}
 \label{fig:struct}
\end{figure} 

We now consider the evolution of the internal structure of the PNS with convection and varying prescriptions for the opacities.  In Fig. \ref{fig:struct}, the evolutions of the entropy and lepton fraction for the two equations of state are shown.  Over the first second  in both models, convection smoothes the entropy and lepton gradients in the outer regions to a state close to neutral buoyancy.  GM3 has a slightly steeper entropy gradient, in part due to its larger $E'_{\rm sym}$, than IU-FSU.  This partially accounts for the slightly larger neutrino luminosity at early times for GM3.  As time progresses, convection steadily digs deeper into the core of the PNS.  For both EoSs, convection proceeds all the way to the core by 15 seconds into the simulation, but it lasts in the interior regions for a much longer period of time for IU-FSU resulting in more rapid lepton depletion in the core.  The exact details of how convection proceeds depend on the initial conditions of the PNS and the behavior of $\gamma_s$ and $\gamma_{Y_L}$ for a given EoS.  For the conditions encountered in the PNS, the variation in $\gamma_{Y_L}$ between the two EoSs we employ is significantly larger than the variation of $\gamma_s$.  More important to the neutrino signal accompanying PNS formation, in GM3 convection ceases in the mantle by $\sim 5$ seconds, whereas convection in the mantle proceeds until $\sim 12$ seconds in IU-FSU.  This difference is mainly driven by the difference in $E'_{\rm sym}$ between the two EoSs.  As the mantle contracts, the second term in Eq.(\ref{eq:dpdyl}) becomes increasingly dominant and is eventually able to stabilize convection.  Qualitatively, increasing $E'_{\rm sym}$ will shut-off convection at an earlier time. 

The depth to which convection penetrates in the core and how long convection proceeds in the core are dependent upon the opacities as well as the EoS.  When only mean field effects on the opacities are considered (i.e. when the neutrino mean free path is shorter), convection does not proceed all the way to the center of the PNS in the GM3 models.  When RPA effects are included, convection does proceed to the central regions of the core.  An increased diffusion rate allows the core to heat up and deleptonize more rapidly, thereby decreasing the stabilizing lepton gradients and increasing the de-stabilizing entropy gradients.

\begin{figure}
\includegraphics[scale=0.3]  {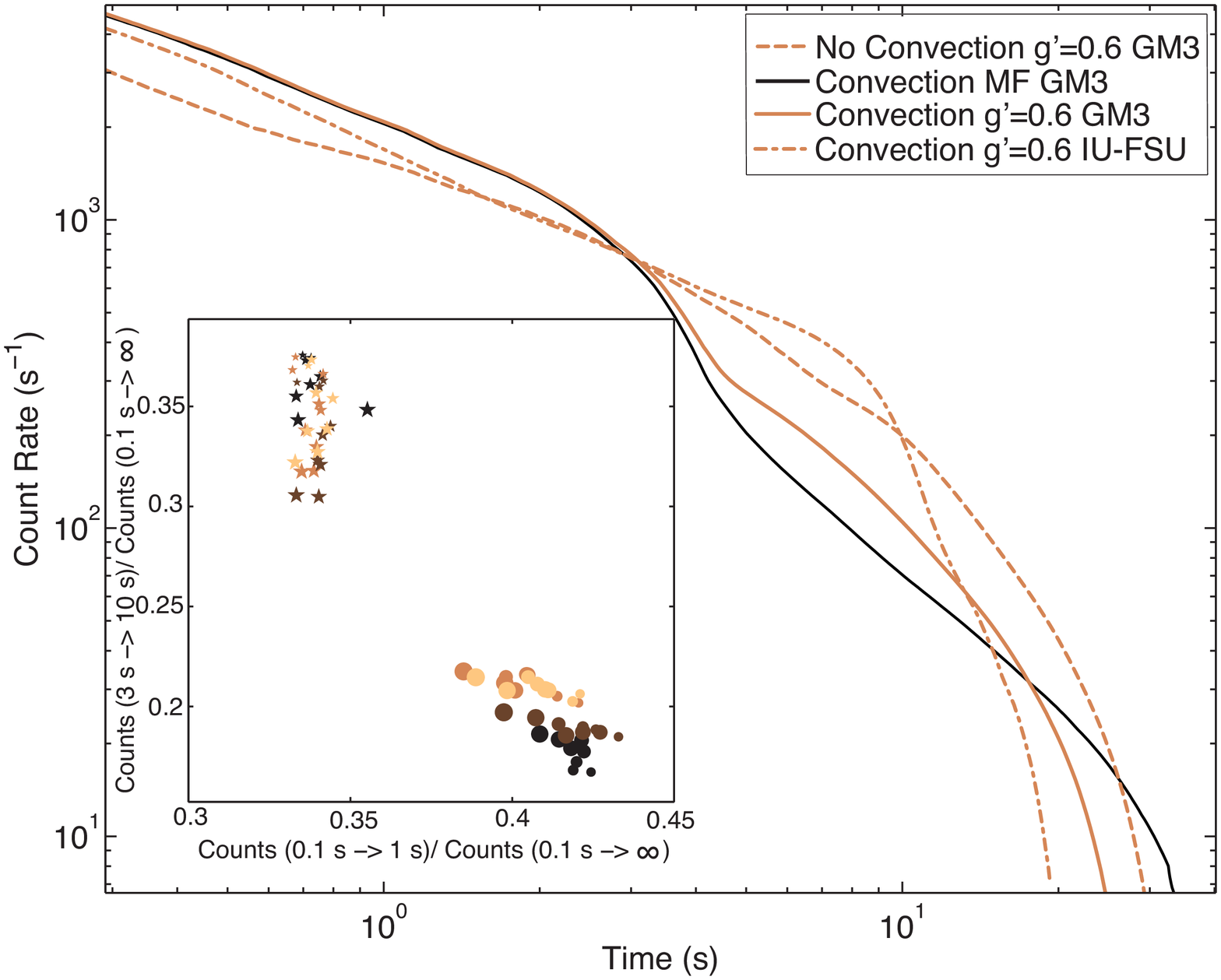}
\caption{Count rates as a function of time for a number of $1.6 M_\odot$ PNS models with and without convection.  The black line is for neutrino opacities calculated in the mean field approximation, while all the other lines are for models that use RPA opacities with $g'=0.6$.  The inset plot shows the integrated number of counts from 0.1 s to 1 s divided by the total number of counts for $t > 0.1$ second on the horizontal axis, and the number of counts for $t > 3$ seconds on divided by the total number of counts for $t > 0.1$ second.  The stars correspond to the IU-FSU EoS and the circles to the GM3 EoS. Symbol sizes correspond to various neutron star rest masses ranging from $1.2 M_\odot$ to $2.1 M_\odot$.  Colors correspond to different values of the Migdal parameter, $g'$.}
 \label{fig:counts}
\end{figure} 

Of course, variations in the convective evolution of the PNS are only interesting to the extent they are potentially observable in the neutrino emission from a nearby supernova.  In Fig. \ref{fig:counts} the expected neutrino count rates for a detector similar to Super Kamiokande-III are shown for a number of PNS cooling models.  We have assumed a threshold energy of 7.5 MeV, a detector mass of 50 kt, a detector efficiency above threshold of unity \cite{Ikeda07}, and a distance of 10 kpc to the supernova.  Equipartition has been assumed between neutrino flavors and the spectral temperatures are calculated from the average radius of neutrino decoupling in the PNS.  The neutrino luminosities and average energies for two models are included as supplemental data.

Both the GM3 and IU-FSU EoSs show enhanced luminosities at early times relative to the models not including convection.  There is only a small difference between the two equations of state at low (sub-nuclear) density, so differences prior to one second are small.  The neutrino count rate is increased by about 30\% relative to the models that do not include convection. This is reasonably consistent with the early time enhancement seen in multi-dimensional models \cite{Buras06b}.  After a second, the count rates between the two EoSs begin to diverge.  The most obvious feature in the count rate for GM3 appears at $\sim 3$ seconds, which is coincident with the end of convection in the mantle.  For the IU-FSU EoS,  the break is also at the time at which mantle convection ends ($\sim 10$ seconds), although it is hard to distinguish from the point at which the PNS becomes optically thin.  As was argued previously, the position of this break reflects the density dependence of the nuclear symmetry energy at $n_B>n_0$ and therefore provides a direct observable of the properties of nuclear matter in the PNS neutrino signal.  Although core convection does not seem to affect the break, it may 
impact the subsequent cooling timescale.

In the inset in Fig. \ref{fig:counts}, integrated neutrino counts over two time windows are shown for a number of PNS masses.  There is a clear separation between the two EoSs independent of mass.  The time of the convective break creates this separation.  This illustrates that this diagnostic of the symmetry energy does not require an accurate determination of the PNS mass.

The inclusion of nucleon correlations through the RPA begins to significantly affect the neutrino emission after about three seconds.  Initially, the luminosities are increased as energy and lepton number are able to more rapidly diffuse out of the core, but at later times the neutrino signal is significantly reduced and drops below the detectable threshold at an earlier time.  

In summary, using a self-consistent model for the PNS core physics, we find that the late time neutrino signal from a core collapse supernova is likely to contain a direct diagnostic of the nuclear symmetry energy at high density.  With current neutrino detectors, these effects should be readily discernible in the neutrino light curve of a single nearby supernova.

\begin{acknowledgements}
Support from the UCOP (09-IR-07-117968-WOOS), the NNSA/DOE SSGF DE-FC52-08NA28752, the US NSF grant AST-0909129, DOE grant DE-AC52-06NA25396 (LANL),  DOE TC on ``Neutrinos and Nucleosynthesis'', and the Spanish MEC grant AYA 2010-21097-C03-02 is gratefully acknowledged.
\end{acknowledgements}
%


\begin{thebibliography}{28}%
\makeatletter
\providecommand \@ifxundefined [1]{%
 \@ifx{#1\undefined}
}%
\providecommand \@ifnum [1]{%
 \ifnum #1\expandafter \@firstoftwo
 \else \expandafter \@secondoftwo
 \fi
}%
\providecommand \@ifx [1]{%
 \ifx #1\expandafter \@firstoftwo
 \else \expandafter \@secondoftwo
 \fi
}%
\providecommand \natexlab [1]{#1}%
\providecommand \enquote  [1]{``#1''}%
\providecommand \bibnamefont  [1]{#1}%
\providecommand \bibfnamefont [1]{#1}%
\providecommand \citenamefont [1]{#1}%
\providecommand \href@noop [0]{\@secondoftwo}%
\providecommand \href [0]{\begingroup \@sanitize@url \@href}%
\providecommand \@href[1]{\@@startlink{#1}\@@href}%
\providecommand \@@href[1]{\endgroup#1\@@endlink}%
\providecommand \@sanitize@url [0]{\catcode `\\12\catcode `\$12\catcode
  `\&12\catcode `\#12\catcode `\^12\catcode `\_12\catcode `\%12\relax}%
\providecommand \@@startlink[1]{}%
\providecommand \@@endlink[0]{}%
\providecommand \url  [0]{\begingroup\@sanitize@url \@url }%
\providecommand \@url [1]{\endgroup\@href {#1}{\urlprefix }}%
\providecommand \urlprefix  [0]{URL }%
\providecommand \Eprint [0]{\href }%
\providecommand \doibase [0]{http://dx.doi.org/}%
\providecommand \selectlanguage [0]{\@gobble}%
\providecommand \bibinfo  [0]{\@secondoftwo}%
\providecommand \bibfield  [0]{\@secondoftwo}%
\providecommand \translation [1]{[#1]}%
\providecommand \BibitemOpen [0]{}%
\providecommand \bibitemStop [0]{}%
\providecommand \bibitemNoStop [0]{.\EOS\space}%
\providecommand \EOS [0]{\spacefactor3000\relax}%
\providecommand \BibitemShut  [1]{\csname bibitem#1\endcsname}%
\let\auto@bib@innerbib\@empty
\bibitem [{\citenamefont {{Burrows}}\ and\ \citenamefont
  {{Lattimer}}(1986)}]{Burrows86}%
  \BibitemOpen
  \bibfield  {author} {\bibinfo {author} {\bibfnamefont {A.}~\bibnamefont
  {{Burrows}}}\ and\ \bibinfo {author} {\bibfnamefont {J.~M.}\ \bibnamefont
  {{Lattimer}}},\ }\href {\doibase 10.1086/164405} {\bibfield  {journal}
  {\bibinfo  {journal} {\apj}\ }\textbf {\bibinfo {volume} {307}},\ \bibinfo
  {pages} {178} (\bibinfo {year} {1986})}\BibitemShut {NoStop}%
\bibitem [{\citenamefont {{Wilson}}\ and\ \citenamefont
  {{Mayle}}(1988)}]{Wilson88}%
  \BibitemOpen
  \bibfield  {author} {\bibinfo {author} {\bibfnamefont {J.~R.}\ \bibnamefont
  {{Wilson}}}\ and\ \bibinfo {author} {\bibfnamefont {R.~W.}\ \bibnamefont
  {{Mayle}}},\ }\href {\doibase 10.1016/0370-1573(88)90036-1} {\bibfield
  {journal} {\bibinfo  {journal} {\physrep}\ }\textbf {\bibinfo {volume}
  {163}},\ \bibinfo {pages} {63} (\bibinfo {year} {1988})}\BibitemShut
  {NoStop}%
\bibitem [{\citenamefont {{Keil}}\ and\ \citenamefont
  {{Janka}}(1995)}]{Keil95a}%
  \BibitemOpen
  \bibfield  {author} {\bibinfo {author} {\bibfnamefont {W.}~\bibnamefont
  {{Keil}}}\ and\ \bibinfo {author} {\bibfnamefont {H.-T.}\ \bibnamefont
  {{Janka}}},\ }\href@noop {} {\bibfield  {journal} {\bibinfo  {journal}
  {\aap}\ }\textbf {\bibinfo {volume} {296}},\ \bibinfo {pages} {145} (\bibinfo
  {year} {1995})}\BibitemShut {NoStop}%
\bibitem [{\citenamefont {{Pons}~{\it et al.}}(1999)}]{Pons99}%
  \BibitemOpen
  \bibfield  {author} {\bibinfo {author} {\bibfnamefont {J.~A.}\ \bibnamefont
  {{Pons}~{\it et al.}}},\ }\href {\doibase 10.1086/306889} {\bibfield
  {journal} {\bibinfo  {journal} {\apj}\ }\textbf {\bibinfo {volume} {513}},\
  \bibinfo {pages} {780} (\bibinfo {year} {1999})}\BibitemShut {NoStop}%
\bibitem [{\citenamefont {{Fischer}~{\it et al.}}(2010)}]{Fischer10}%
  \BibitemOpen
  \bibfield  {author} {\bibinfo {author} {\bibfnamefont {T.}~\bibnamefont
  {{Fischer}~{\it et al.}}},\ }\href {\doibase 10.1051/0004-6361/200913106}
  {\bibfield  {journal} {\bibinfo  {journal} {\aap}\ }\textbf {\bibinfo
  {volume} {517}},\ \bibinfo {pages} {A80+} (\bibinfo {year}
  {2010})}\BibitemShut {NoStop}%
\bibitem [{\citenamefont {{H{\"u}depohl}~{\it et al.}}(2010)}]{Huedepohl10}%
  \BibitemOpen
  \bibfield  {author} {\bibinfo {author} {\bibfnamefont {L.}~\bibnamefont
  {{H{\"u}depohl}~{\it et al.}}},\ }\href {\doibase
  10.1103/PhysRevLett.104.251101} {\bibfield  {journal} {\bibinfo  {journal}
  {Physi. Rev. Lett.}\ }\textbf {\bibinfo {volume} {104}},\ \bibinfo {pages}
  {251101} (\bibinfo {year} {2010})}\BibitemShut {NoStop}%
\bibitem [{\citenamefont {{Keil}}\ \emph {et~al.}(199)\citenamefont {{Keil}},
  \citenamefont {{Janka}},\ and\ \citenamefont {{Raffelt}}}]{Keil95b}%
  \BibitemOpen
  \bibfield  {author} {\bibinfo {author} {\bibfnamefont {W.}~\bibnamefont
  {{Keil}}}, \bibinfo {author} {\bibfnamefont {H.-T.}\ \bibnamefont {{Janka}}},
  \ and\ \bibinfo {author} {\bibfnamefont {G.}~\bibnamefont {{Raffelt}}},\
  }\href@noop {} {\bibfield  {journal} {\bibinfo  {journal} {\prd}\ }\textbf
  {\bibinfo {volume} {51}},\ \bibinfo {pages} {6635} (\bibinfo {year}
  {1995})}\BibitemShut {NoStop}%
\bibitem [{\citenamefont {{Pons}~{\it et al.}}(2001{\natexlab{a}})}]{Pons01a}%
  \BibitemOpen
  \bibfield  {author} {\bibinfo {author} {\bibfnamefont {J.~A.}\ \bibnamefont
  {{Pons}~{\it et al.}}},\ }\href {\doibase 10.1086/320642} {\bibfield
  {journal} {\bibinfo  {journal} {\apj}\ }\textbf {\bibinfo {volume} {553}},\
  \bibinfo {pages} {382} (\bibinfo {year} {2001}{\natexlab{a}})}\BibitemShut
  {NoStop}%
\bibitem [{\citenamefont {{Pons}~{\it et al.}}(2001{\natexlab{b}})}]{Pons01b}%
  \BibitemOpen
  \bibfield  {author} {\bibinfo {author} {\bibfnamefont {J.~A.}\ \bibnamefont
  {{Pons}~{\it et al.}}},\ }\href {\doibase 10.1103/PhysRevLett.86.5223}
  {\bibfield  {journal} {\bibinfo  {journal} {Phys. Rev. Lett.}\ }\textbf
  {\bibinfo {volume} {86}},\ \bibinfo {pages} {5223} (\bibinfo {year}
  {2001}{\natexlab{b}})}\BibitemShut {NoStop}%
\bibitem [{\citenamefont {{Epstein}}(1979)}]{Epstein79}%
  \BibitemOpen
  \bibfield  {author} {\bibinfo {author} {\bibfnamefont {R.~I.}\ \bibnamefont
  {{Epstein}}},\ }\href@noop {} {\bibfield  {journal} {\bibinfo  {journal}
  {\mnras}\ }\textbf {\bibinfo {volume} {188}},\ \bibinfo {pages} {305}
  (\bibinfo {year} {1979})}\BibitemShut {NoStop}%
\bibitem [{\citenamefont {{Burrows}}(1987)}]{Burrows87}%
  \BibitemOpen
  \bibfield  {author} {\bibinfo {author} {\bibfnamefont {A.}~\bibnamefont
  {{Burrows}}},\ }\href {\doibase 10.1086/184937} {\bibfield  {journal}
  {\bibinfo  {journal} {\apjl}\ }\textbf {\bibinfo {volume} {318}},\ \bibinfo
  {pages} {L57} (\bibinfo {year} {1987})}\BibitemShut {NoStop}%
\bibitem [{\citenamefont {{Keil}}\ \emph {et~al.}(1996)\citenamefont {{Keil}},
  \citenamefont {{Janka}},\ and\ \citenamefont {{Mueller}}}]{Keil96}%
  \BibitemOpen
  \bibfield  {author} {\bibinfo {author} {\bibfnamefont {W.}~\bibnamefont
  {{Keil}}}, \bibinfo {author} {\bibfnamefont {H.}~\bibnamefont {{Janka}}}, \
  and\ \bibinfo {author} {\bibfnamefont {E.}~\bibnamefont {{Mueller}}},\
  }\href@noop {} {\bibfield  {journal} {\bibinfo  {journal} {\apjl}\ }\textbf
  {\bibinfo {volume} {473}},\ \bibinfo {pages} {L111+} (\bibinfo {year}
  {1996})}\BibitemShut {NoStop}%
\bibitem [{\citenamefont {{Mezzacappa}~{\it et al.}}(1998)}]{Mezzacappa98a}%
  \BibitemOpen
  \bibfield  {author} {\bibinfo {author} {\bibfnamefont {A.}~\bibnamefont
  {{Mezzacappa}~{\it et al.}}},\ }\href {\doibase 10.1086/305164} {\bibfield
  {journal} {\bibinfo  {journal} {\apj}\ }\textbf {\bibinfo {volume} {493}},\
  \bibinfo {pages} {848} (\bibinfo {year} {1998})}\BibitemShut {NoStop}%
\bibitem [{\citenamefont {{Dessart}~{\it et al.}}(2006)}]{Dessart06}%
  \BibitemOpen
  \bibfield  {author} {\bibinfo {author} {\bibfnamefont {L.}~\bibnamefont
  {{Dessart}~{\it et al.}}},\ }\href {\doibase 10.1086/504068} {\bibfield
  {journal} {\bibinfo  {journal} {\apj}\ }\textbf {\bibinfo {volume} {645}},\
  \bibinfo {pages} {534} (\bibinfo {year} {2006})}\BibitemShut {NoStop}%
\bibitem [{\citenamefont {{Buras}~{\it et al.}}(2006)}]{Buras06b}%
  \BibitemOpen
  \bibfield  {author} {\bibinfo {author} {\bibfnamefont {R.}~\bibnamefont
  {{Buras}~{\it et al.}}},\ }\href {\doibase 10.1051/0004-6361:20054654}
  {\bibfield  {journal} {\bibinfo  {journal} {\aap}\ }\textbf {\bibinfo
  {volume} {457}},\ \bibinfo {pages} {281} (\bibinfo {year}
  {2006})}\BibitemShut {NoStop}%
\bibitem [{\citenamefont {{Bruenn}}\ and\ \citenamefont
  {{Dineva}}(1996)}]{Bruenn96}%
  \BibitemOpen
  \bibfield  {author} {\bibinfo {author} {\bibfnamefont {S.~W.}\ \bibnamefont
  {{Bruenn}}}\ and\ \bibinfo {author} {\bibfnamefont {T.}~\bibnamefont
  {{Dineva}}},\ }\href {\doibase 10.1086/309921} {\bibfield  {journal}
  {\bibinfo  {journal} {\apjl}\ }\textbf {\bibinfo {volume} {458}},\ \bibinfo
  {pages} {L71+} (\bibinfo {year} {1996})}\BibitemShut {NoStop}%
\bibitem [{\citenamefont {{Miralles}}\ \emph {et~al.}(2000)\citenamefont
  {{Miralles}}, \citenamefont {{Pons}},\ and\ \citenamefont
  {{Urpin}}}]{Miralles00}%
  \BibitemOpen
  \bibfield  {author} {\bibinfo {author} {\bibfnamefont {J.~A.}\ \bibnamefont
  {{Miralles}}}, \bibinfo {author} {\bibfnamefont {J.~A.}\ \bibnamefont
  {{Pons}}}, \ and\ \bibinfo {author} {\bibfnamefont {V.~A.}\ \bibnamefont
  {{Urpin}}},\ }\href {\doibase 10.1086/317163} {\bibfield  {journal} {\bibinfo
   {journal} {\apj}\ }\textbf {\bibinfo {volume} {543}},\ \bibinfo {pages}
  {1001} (\bibinfo {year} {2000})}\BibitemShut {NoStop}%
\bibitem [{\citenamefont {{Tsang}~{\it et al.}}(2011)}]{Tsang11}%
  \BibitemOpen
  \bibfield  {author} {\bibinfo {author} {\bibfnamefont {M.~B.}\ \bibnamefont
  {{Tsang}~{\it et al.}}},\ }\href {\doibase 10.1016/j.ppnp.2011.01.041}
  {\bibfield  {journal} {\bibinfo  {journal} {Prog. Part. Nucl. Phys.}\
  }\textbf {\bibinfo {volume} {66}},\ \bibinfo {pages} {400} (\bibinfo {year}
  {2011})}\BibitemShut {NoStop}%
\bibitem [{\citenamefont {{Fattoyev}~{\it et al.}}(2010)}]{Fattoyev10}%
  \BibitemOpen
  \bibfield  {author} {\bibinfo {author} {\bibfnamefont {F.~J.}\ \bibnamefont
  {{Fattoyev}~{\it et al.}}},\ }\href {\doibase 10.1103/PhysRevC.82.055803}
  {\bibfield  {journal} {\bibinfo  {journal} {\prc}\ }\textbf {\bibinfo
  {volume} {82}},\ \bibinfo {pages} {055803} (\bibinfo {year}
  {2010})}\BibitemShut {NoStop}%
\bibitem [{\citenamefont {{Gandolfi}}\ \emph {et~al.}(2011)\citenamefont
  {{Gandolfi}}, \citenamefont {{Carlson}},\ and\ \citenamefont
  {{Reddy}}}]{Gandolfi11}%
  \BibitemOpen
  \bibfield  {author} {\bibinfo {author} {\bibfnamefont {S.}~\bibnamefont
  {{Gandolfi}}}, \bibinfo {author} {\bibfnamefont {J.}~\bibnamefont
  {{Carlson}}}, \ and\ \bibinfo {author} {\bibfnamefont {S.}~\bibnamefont
  {{Reddy}}},\ }\href@noop {} {\bibfield  {journal} {\bibinfo  {journal} {ArXiv
  e-prints}\ } (\bibinfo {year} {2011})},\ \Eprint
  {http://arxiv.org/abs/1101.1921} {arXiv:1101.1921} \BibitemShut
  {NoStop}%
\bibitem [{\citenamefont {{Glendenning}}\ and\ \citenamefont
  {{Moszkowski}}(1991)}]{Glendenning91}%
  \BibitemOpen
  \bibfield  {author} {\bibinfo {author} {\bibfnamefont {N.~K.}\ \bibnamefont
  {{Glendenning}}}\ and\ \bibinfo {author} {\bibfnamefont {S.~A.}\ \bibnamefont
  {{Moszkowski}}},\ }\href {\doibase 10.1103/PhysRevLett.67.2414} {\bibfield
  {journal} {\bibinfo  {journal} {Phys. Rev. Lett.}\ }\textbf {\bibinfo
  {volume} {67}},\ \bibinfo {pages} {2414} (\bibinfo {year}
  {1991})}\BibitemShut {NoStop}%
\bibitem [{\citenamefont {{Steiner}~{\it et al.}}(2005)}]{Steiner05}%
  \BibitemOpen
  \bibfield  {author} {\bibinfo {author} {\bibfnamefont {A.~W.}\ \bibnamefont
  {{Steiner}~{\it et al.}}},\ }\href {\doibase 10.1016/j.physrep.2005.02.004}
  {\bibfield  {journal} {\bibinfo  {journal} {Phys. Rep.}\ }\textbf {\bibinfo
  {volume} {411}},\ \bibinfo {pages} {325} (\bibinfo {year}
  {2005})}\BibitemShut {NoStop}%
\bibitem [{\citenamefont {{Lattimer}}\ and\ \citenamefont
  {{Prakash}}(2001)}]{Lattimer01}%
  \BibitemOpen
  \bibfield  {author} {\bibinfo {author} {\bibfnamefont {J.~M.}\ \bibnamefont
  {{Lattimer}}}\ and\ \bibinfo {author} {\bibfnamefont {M.}~\bibnamefont
  {{Prakash}}},\ }\href {\doibase 10.1086/319702} {\bibfield  {journal}
  {\bibinfo  {journal} {\apj}\ }\textbf {\bibinfo {volume} {550}},\ \bibinfo
  {pages} {426} (\bibinfo {year} {2001})}\BibitemShut {NoStop}%
\bibitem [{\citenamefont {Horowitz}\ and\ \citenamefont
  {Piekarewicz}(2001)}]{Horowitz01}%
  \BibitemOpen
  \bibfield  {author} {\bibinfo {author} {\bibfnamefont {C.~J.}\ \bibnamefont
  {Horowitz}}\ and\ \bibinfo {author} {\bibfnamefont {J.}~\bibnamefont
  {Piekarewicz}},\ }\href@noop {} {\bibfield  {journal} {\bibinfo  {journal}
  {Phys. Rev. C}\ }\textbf {\bibinfo {volume} {64}},\ \bibinfo {pages} {062802}
  (\bibinfo {year} {2001})}\BibitemShut {NoStop}%
\bibitem [{\citenamefont {{Reddy}~{\it et al.}}(1999)}]{Reddy99}%
  \BibitemOpen
  \bibfield  {author} {\bibinfo {author} {\bibfnamefont {S.}~\bibnamefont
  {{Reddy}~{\it et al.}}},\ }\href {\doibase 10.1103/PhysRevC.59.2888}
  {\bibfield  {journal} {\bibinfo  {journal} {\prc}\ }\textbf {\bibinfo
  {volume} {59}},\ \bibinfo {pages} {2888} (\bibinfo {year}
  {1999})}\BibitemShut {NoStop}%
\bibitem [{\citenamefont {{Kim}}\ \emph {et~al.}(1995)\citenamefont {{Kim}},
  \citenamefont {{Piekarewicz}},\ and\ \citenamefont {{Horowitz}}}]{Kim95}%
  \BibitemOpen
  \bibfield  {author} {\bibinfo {author} {\bibfnamefont {H.}~\bibnamefont
  {{Kim}}}, \bibinfo {author} {\bibfnamefont {J.}~\bibnamefont
  {{Piekarewicz}}}, \ and\ \bibinfo {author} {\bibfnamefont {C.~J.}\
  \bibnamefont {{Horowitz}}},\ }\href {\doibase 10.1103/PhysRevC.51.2739}
  {\bibfield  {journal} {\bibinfo  {journal} {\prc}\ }\textbf {\bibinfo
  {volume} {51}},\ \bibinfo {pages} {2739} (\bibinfo {year}
  {1995})}\BibitemShut {NoStop}%
\bibitem [{\citenamefont {Fantoni}\ \emph {et~al.}(2001)\citenamefont
  {Fantoni}, \citenamefont {Sarsa},\ and\ \citenamefont {Schmidt}}]{Fantoni01}%
  \BibitemOpen
  \bibfield  {author} {\bibinfo {author} {\bibfnamefont {S.}~\bibnamefont
  {Fantoni}}, \bibinfo {author} {\bibfnamefont {A.}~\bibnamefont {Sarsa}}, \
  and\ \bibinfo {author} {\bibfnamefont {K.~E.}\ \bibnamefont {Schmidt}},\
  }\href@noop {} {\bibfield  {journal} {\bibinfo  {journal} {Phys. Rev. Lett.}\
  }\textbf {\bibinfo {volume} {87}},\ \bibinfo {pages} {181101} (\bibinfo
  {year} {2001})}\BibitemShut {NoStop}%
\bibitem [{\citenamefont {{Ikeda}~{\it et al.}}(2007)}]{Ikeda07}%
  \BibitemOpen
  \bibfield  {author} {\bibinfo {author} {\bibfnamefont {M.}~\bibnamefont
  {{Ikeda}~{\it et al.}}},\ }\href {\doibase 10.1086/521547} {\bibfield
  {journal} {\bibinfo  {journal} {\apj}\ }\textbf {\bibinfo {volume} {669}},\
  \bibinfo {pages} {519} (\bibinfo {year} {2007})}\BibitemShut {NoStop}%
\end{thebibliography}
\end{document}